\documentstyle[pre,aps,epsf,floats]{revtex} 

\begin{document} 

\def\bottomfraction{0.5}
\textheight24.6cm
\flushbottom

\twocolumn[\hsize\textwidth\columnwidth\hsize\csname @twocolumnfalse\endcsname


\title{Scaling behavior of the absorbing phase transition\\
in a conserved lattice gas around the upper critical dimension}

\author{S. L\"ubeck}

\address{
Theoretische Tieftemperaturphysik, 
Gerhard-Mercator-Universit\"at Duisburg,\\ 
Lotharstr. 1, 47048 Duisburg, Germany \\}

\date{Received March 1, 2001}

\maketitle

\begin{abstract}
We analyse numerically the critical behavior of a
conserved lattice gas which was recently introduced 
as an example of the new universality class of 
absorbing phase transitions with a conserved field  
\protect{[}Phys.~Rev.~Lett.~85, 1803 (2000)\protect{]}.
We determine the critical exponent of the order parameter as well 
as the critical exponent of the 
order parameter fluctuations in $D=2,3,4,5$ dimensions.
A comparison of our results and those
obtained from a mean-field approach 
and a field theory
suggests that the upper critical dimension
of the absorbing phase transition is four.
\end{abstract}

\pacs{05.70.Ln, 05.50.+q, 05.65.+b}

]  

\setcounter{page}{1}
\markright{{\it Physical Review E} {\bf 64}, 016123 (2001)}
\thispagestyle{myheadings}
\pagestyle{myheadings}

\section{Introduction}

The scaling behavior of directed percolation is recognized 
as the paradigmatic example of the critical behavior
of several non-equilibrium systems which exhibits a 
continuous phase transition from an active state
to an absorbing non-active state 
(see for instance~\cite{GRINSTEIN_2,HINRICHSEN_1}).
Such systems are common in physics, biology, as well 
as catalytic chemical reactions.
This widespread occurrence 
corresponds to the well known universality
hypothesis of Janssen and Grassberger that models which 
exhibit a continuous phase transition to a single
absorbing state generally belong to the universality class
of directed percolation~\cite{JANSSEN_1,GRASSBERGER_2}.

Recently Rossi {\it et at.}~introduced a conserved
lattice gas (CLG) with a stochastic short range 
interaction that exhibits a continuous phase transition
to an absorbing state at a critical value of the
particle density~\cite{ROSSI_1}.
The CLG model is expected to belong to a new 
universality class of absorbing phase transitions 
characterized by a conserved field.
Similar to the above hypothesis the authors 
conjectured that "all stochastic models with an 
infinite number of absorbing states in which the order
parameter evolution is coupled to a nondiffusive
conserved field  define a unique universality 
class"~\cite{ROSSI_1}.
Besides the CLG model the authors considered 
the conserved threshold transfer process model 
as well as a modification of the stochastic 
sandpile model of Manna~\cite{MANNA_2} 
and observed numerically compatible values of 
the critical exponents.
Furthermore a reaction-diffusion model was 
introduced in~\cite{PASTOR_2} which is expected
to belong to the same universality class.
This reaction diffusion model allows to derive 
a field theoretical description 
which is expected to represent the critical 
behavior of the whole universality class~\cite{PASTOR_2}.

In this work we consider for the first time
the scaling behavior of the CLG model and 
therefore the critical behavior of the 
new universality class in higher dimensions.
We determine numerically the critical
exponent of the order parameter as well as
the exponent of the order parameter fluctuations.
Our results show that the values of the exponents 
depend on the dimension for $D\le 4$.
Above this dimension we observe a 
mean-field scaling behavior.
Thus our results suggest that the upper critical
dimension of the CLG model is four
as already predicted from the field theoretical
approach~\cite{PASTOR_2}.

\section{d=2}
\label{sec:d2}

We consider the CLG model on $D$-dimensional cubic
lattices of linear size~$L$.
Initially one distributes randomly $N=\rho L$ 
particles on the system where $\rho$ denotes the
particle density.
In order to mimic a repulsive interaction
a given particle is considered as {\it active} if at least one of its 
$2D$ neighboring sites on the cubic lattice is occupied
by another particle.
If all neighboring sites are empty the particle
remains {\it inactive}.
Active particles are moved in the next update step to one 
of their empty nearest neighbor sites, selected at random.
Starting from a random distribution of particles
the system reach after a transient regime a
steady state which is characterized by the
density of active sites $\rho_{\rm a}$.
The density $\rho_{\rm a}$ is the order parameter of 
the absorbing phase transition, i.e., it vanishes 
if the control parameter $\rho$ is lower
than the critical value $\rho_{\rm c}$.
In contrast to the work of Rossi {\it et al.}~\cite{ROSSI_1},
who used a parallel update scheme, 
we applied in our simulations a random sequential 
update, i.e., all active sites are listed, and then updated 
in a randomly chosen sequence.

\begin{figure}[t]
 \epsfxsize=7.1cm
 \epsffile{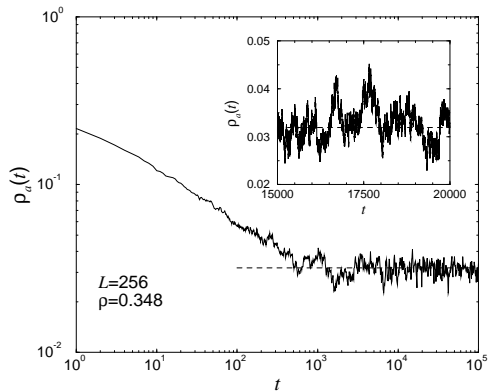}
  \caption{
    The density of active sites $\rho_{\rm a}$ as a function
    of time (number of update steps) for a certain value of $\rho$.
    After a transient regime, which depends on the initial configuration,
    the density of active sites
    fluctuates around the steady state value $\langle \rho_{\rm a} \rangle$
    (dashed line).
   }
 \label{fig:rho_a_t_01} 
\end{figure}

We consider in two dimensions simple cubic systems of linear 
size $L=32,64,128,\ldots, 2048$ 
with periodic boundary conditions.
Starting with a random configuration of particles,
a sufficient number of update steps 
has to be performed to reach the steady state
where the number of active sites fluctuates around 
an average value (see Fig.\,\ref{fig:rho_a_t_01}).
Approaching the transition point, more and more update
steps are needed to reach this steady state.
For instance in the case $L=2048$ we use
$2\,10^6$ update steps to "equilibrate" the system.
In the steady state the number of active sites is monitored for 
$5\,10^5$ update steps.
This procedure is repeated in all dimensions 
for at least $10$ different initial configurations.
From this data we determine the average density
of active sites $\langle \rho_{\rm a} \rangle$ as well
as its fluctuations
\begin{equation}
\Delta \rho_{\rm a} \; = \; L^D \, \left (\langle \rho_{\rm a}^2 \rangle
\, - \, \langle \rho_{\rm a}\rangle^2 \right ).
\label{eq:fluc_01}
\end{equation}

\begin{figure}[b]
 \epsfxsize=7.1cm
 \epsffile{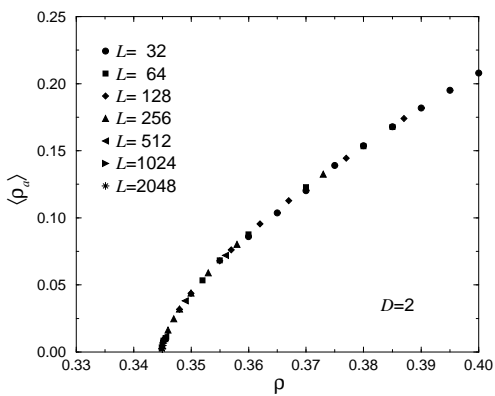}
  \caption{
    The average density of active sites $\langle \rho_{\rm a} \rangle$
    as a function of the global particle density $\rho$
    for various system sizes~$L$ in $D=2$ dimensions.
   }
 \label{fig:rho_a_2d_01} 
\end{figure}

\begin{figure}[t]
 \epsfxsize=7.1cm
 \epsffile{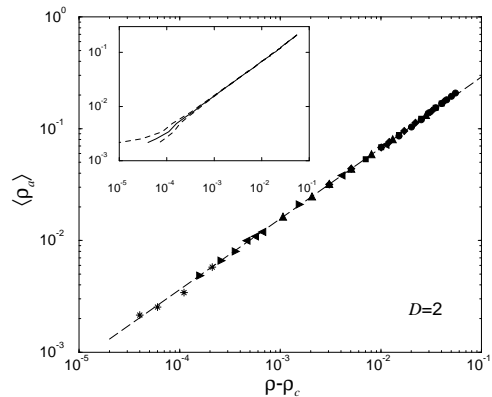}
  \caption{
    The average density of active sites $\langle \rho_{\rm a} \rangle$
    as a function of $\rho-\rho_{\rm c}$ in $D=2$.
    The symbols mark different system sizes~$L$
    (see Fig.\,\protect\ref{fig:rho_a_2d_01}). 
    The dashed line corresponds to a power-law fit with
    $\rho_{\rm c}=0.34494\pm0.00003$ and $\beta=0.637\pm0.009$.
    In the inset we display the same data for 
    $\rho_{\rm c}=0.34491$ and $\rho_{\rm c}=0.34497$.
    Compared to the above value (solid line) both curves
    are characterized by significant curvatures in the 
    plotted log-log diagram.
    For the sake of simplicity we plot in the inset lines
    instead of symbols.
   }
 \label{fig:rho_a_2d_02} 
\end{figure}

In Fig.~\ref{fig:rho_a_2d_01} we plot the average
density of active sites as a function of the 
particle density~$\rho$ for various system sizes~$L$.
As one can see $\langle \rho_{\rm a} \rangle$ tends
to zero in the vicinity of $\rho\approx 0.345$.
Assuming that the scaling behavior of the density of 
active sites is given by
\begin{equation}
\langle \rho_{\rm a} \rangle \; \sim \; (\rho - \rho_{\rm c})^\beta 
\label{eq:order_par_01}
\end{equation}
one varies $\rho_{\rm c}$ until one gets a straight line
in a log-log plot.
Convincing results are obtained for $\rho_{\rm c}=0.34494\pm0.00003$
and the corresponding curve is shown in
Fig.\,\ref{fig:rho_a_2d_02}.
For $\rho_{\rm c}=0.34491$ and $\rho_{\rm c}=0.34497$ we observe
significant curvatures in the log-log plot
(see inset of Fig.\,\ref{fig:rho_a_2d_02}).
In this way we estimate the error-bars in the determination
of the critical density.
A regression analysis yields the value
of the order parameter exponent $\beta=0.637\pm0.009$.
Rossi {\it et al.}~reported the values $\beta=0.63\pm 0.01$
and $\rho_{\rm c}=0.28875$, obtained from simulations
with a parallel update scheme and of smaller system sizes 
($L\le 512$)~\cite{ROSSI_1}.
Thus we see that the different update scheme
affects only the value of the critical density 
but not the critical exponent $\beta$.
Furthermore the value of $\beta$ differs from
the corresponding value of the directed percolation
universality class $\beta_{\scriptscriptstyle \rm DP}=0.584\pm0.004$ 
(see \cite{HINRICHSEN_1}).

\begin{figure}[b]
 \epsfxsize=7.1cm
 \epsffile{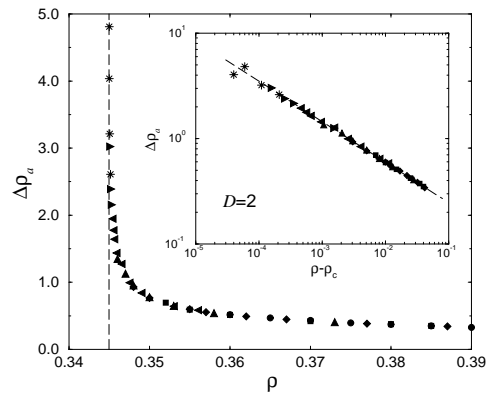}
  \caption{
     The fluctuations of the order parameter $\Delta \rho_{\rm a}$
     as a function of the global density $\rho$ in $D=2$.
     The symbols mark different system sizes 
     (see Fig.\,\protect\ref{fig:rho_a_2d_01}) and the 
     dashed line corresponds to the critical density
     $\rho_{\rm c} = 0.34494$.
     The inset displays the fluctuations as a function of
     $\rho - \rho_{\rm c}$.
     The dashed line corresponds to a power-law behavior
     [Eq.\,(\protect\ref{eq:fluc_02})] with an exponent
     $\gamma=0.384\pm0.023$.
   }
 \label{fig:rho_a_2d_03} 
\end{figure}

The fluctuations of the order parameter [Eq.\,(\ref{eq:fluc_01})]
are plotted in Fig.\,\ref{fig:rho_a_2d_03}.
Approaching the transition point $\Delta \rho_{\rm a}$
increases and diverges at $\rho_{\rm c}$.
Close to the critical point the fluctuations 
scale as
\begin{equation}
\Delta \rho_{\rm a} \; \sim \; (\rho - \rho_{\rm c})^{-\gamma} 
\label{eq:fluc_02}
\end{equation}
in the active phase (see inset of Fig.\,\ref{fig:rho_a_2d_03}).
Using a regression analysis one gets $\gamma=0.384\pm0.023$.

\begin{figure}[t]
 \epsfxsize=7.1cm
 \epsffile{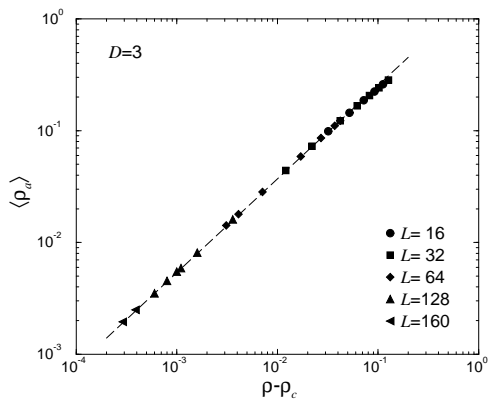}
  \caption{
    The average density of active sites $\langle \rho_{\rm a} \rangle$
    as a function of $\rho-\rho_{\rm c}$ for various system sizes~$L$
    in $D=3$.
    The dashed line corresponds to a power-law fit with
     $\rho_{\rm c}=0.2179\pm0.0001$ and $\beta=0.837\pm0.015$.
   }
 \label{fig:rho_a_3d_02} 
\end{figure}

\section{d=3}
\label{sec:d3}

\begin{figure}[b]
 \epsfxsize=7.1cm
 \epsffile{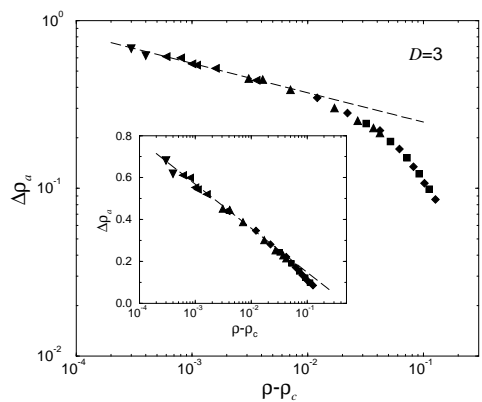}
  \caption{
     The fluctuations of the order parameter $\Delta \rho_{\rm a}$
     as a function of $\rho-\rho_{\rm c}$ in $D=3$.
     The symbols mark different system sizes 
     (see Fig.\,\protect\ref{fig:rho_a_3d_02}).
     The data can be interpreted either as a power-law
     with a small exponent $\gamma$ 
     (the dashed line corresponds to $\gamma=0.18\pm0.06$) 
     or as a logarithmic
     growth [see Eq.\,(\protect\ref{eq:fluc_3d_log}) and inset, where
     the dashed line is to guide the eye].
   }
 \label{fig:rho_a_3d_03} 
\end{figure}

For the three dimensional
model system sizes from $L=16$ to $L=160$ are considered.
Close to the transition point $5\,10^6$ update steps are used 
to reach the steady state in the case of the
largest system size.
The obtained results for the density of active
sites $\langle  \rho_{\rm a} \rangle$ are 
shown in Fig.\,\ref{fig:rho_a_3d_02}. 
A straight line in a log-log plot is obtained
for $\rho_{\rm c}=0.2179\pm0.0001$ and the
corresponding value of the order parameter
exponent is $\beta=0.837\pm0.015$.
The error-bars of the critical density are determined
in the same way as for $D=2$.
The value of the order parameter exponent agrees
within the error-bars with the corresponding
values of the Manna sandpile model ($\beta=0.84\pm0.02$)
and of a reaction diffusion model ($\beta=0.86\pm0.02$) 
which are expected to belong to the same universality
class~\cite{PASTOR_2}.
Our result differs slightly from the corresponding value
of the directed percolation universality class
$\beta_{\scriptscriptstyle \rm DP}=0.81\pm0.01$
(see \cite{HINRICHSEN_1}).

Similar to the two dimensional case the
order parameter fluctuations $\Delta \rho_{\rm a}$  
display a maximum at the transition point
but the dependence of $\Delta \rho_{\rm a}$ on the
density of particles is not clear.
In Fig.\,\ref{fig:rho_a_3d_03} we plot the fluctuations
in a log-log plot.
It seems that the asymptotic behavior corresponds
to a power-law [Eq.\,(\ref{eq:fluc_02})]
with an exponent $\gamma=0.18\pm 0.06$.
But as the inset of Fig.\,\ref{fig:rho_a_3d_03}
shows the data are also consistent with the
assumption that the fluctuations are characterized
by a logarithmic divergence (i.e.~$\gamma=0$) 
at $\rho_{\rm c}$ according to
\begin{equation}
\Delta \rho_{\rm a } \; \sim \;
| \ln{(\rho-\rho_{\rm c})}|.
\label{eq:fluc_3d_log}
\end{equation}
Thus the numerical data indicate that the
fluctuations $\Delta \rho_{\rm a}$ diverge at the
critical point but the data can be interpreted either
as a power-law with a small exponent or as a
logarithmic growth.
Further investigations are needed to clarify
this point.

\section{d=4}
\label{sec:d4}

\begin{figure}[b]
 \epsfxsize=7.1cm
 \epsffile{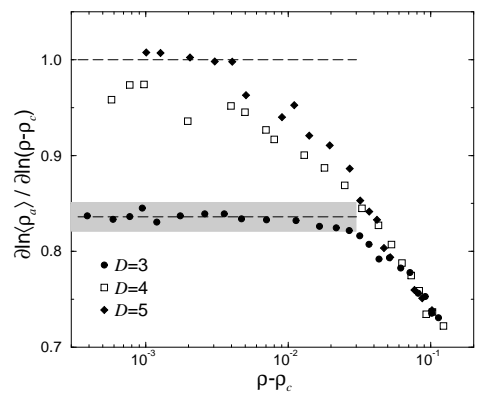}
  \caption{
    The logarithmic derivative [Eq.\,(\protect\ref{eq:log_derivative})]
    as a function of $\rho-\rho_{\rm c}$. 
    The logarithmic derivative can be interpreted as an
    effective exponent $\beta_{\rm eff}$. 
    The figure shows that the four-dimensional
    exponent does not display a saturation as the
    exponents of the three and five dimensional model do.
    The dashed lines correspond to the three dimensional
    value $\beta=0.837\pm0.015$ and the mean-field value $\beta=1$,
    respectively.
    The shadowed region marks the uncertainty of the
    determination of $\beta$.
   }
 \label{fig:log_diff} 
\end{figure}

In order to analyze the scaling behavior of the
four dimensional CLG model we performed
numerical simulations with system sizes $L\in\{8,16,32,48\}$.
In the case of the largest system size $6\, 10^6$ updates steps
were used to reach the steady state.
Plotting the values of the average particle 
density $\langle \rho_{\rm a} \rangle$ in a log-log
plot no straight line, i.e., no pure power-law
behavior could be observed.
To illustrate this behavior we plot in 
Fig.\,\ref{fig:log_diff} the logarithmic derivative
\begin{equation}
\frac{\partial\,\ln{\langle \rho_{\rm a} \rangle}\;\;}
{\partial\,\ln{(\rho-\rho_{\rm c})}},
\label{eq:log_derivative}
\end{equation}
which can be interpreted as an effective 
exponent~$\beta_{\rm eff}$.
If the scaling behavior of the active site
density is given by Eq.\,(\ref{eq:order_par_01})
the logarithmic derivative tends to the value of
$\beta$ for $\rho-\rho_{\rm c}\to 0$.
This behavior is observed in the three dimensional
case (see Fig.\,\ref{fig:log_diff}).
For $D=4$ the logarithmic derivative displays
no saturation for $\rho\to \rho_{\rm c}$, i.e.,
the scaling behavior of the four dimensional
model can not be described by a simple
power-law behavior. 
Significant corrections to the usual scaling
behavior [Eq.\,(\ref{eq:order_par_01})] 
occur for instance at the upper critical
dimensional where the scaling behavior is
governed by the mean-field exponents modified
by logarithmic corrections.

\begin{figure}[t]
 \epsfxsize=7.1cm
 \epsffile{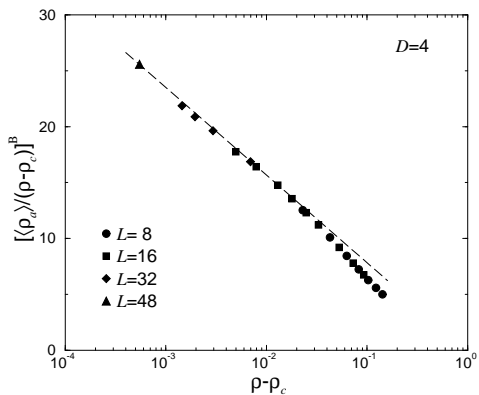}
  \caption{
     The density of active sites in $D=4$ rescaled 
     according to Eq.\,(\protect\ref{eq:order_par_dc}).
     The assumed asymptotic behavior (straight line in
     a log-log plot) is obtained for
     $\rho_{\rm c}=0.1571 \pm0.0002$ and ${\rm B}=0.39$.
     Thus the scaling behavior of the order parameter
     of the four dimensional model is governed by the
     mean-field exponent $\beta=1$ modified by logarithmic
     corrections.
     The dashed line is just to guide the eye.
   }
 \label{fig:rho_a_4d_02} 
\end{figure}

Recently a modified version of the 
CLG model was introduced where the active 
particles are distributed to randomly chosen 
empty lattice sites~\cite{LUEB_20}.
Since the randomness of the particle hopping breaks 
long range spatial correlations this model is expected to
be characterized by the mean-field scaling
behavior of the CLG model.
Mapping the dynamics of this random hopping CLG model
to a simple branching process one can derive the
critical exponent $\beta=1$.
This value of the order parameter exponent is also 
obtained from a field theoretical description 
of the CLG model~\cite{PASTOR_4} 
and was already predicted from a phenomenological 
field theory~\cite{VESPIGNANI_4} 
of the so-called fix-energy stochastic sandpile model 
which is expected to be in the same universality class.

Thus we assume that the scaling behavior 
of the order parameter is given in leading order 
by the ansatz
\begin{equation}
\langle \rho_{\rm a} \rangle \; \sim \; 
(\rho - \rho_{\rm c})^\beta \;
| \ln{(\rho-\rho_{\rm c})}|^{\rm B}
\label{eq:order_par_dc}
\end{equation}
with $\beta=1$.
Therefore we varied in our analysis $\rm B$ and
$\rho_{\rm c}$ until we get the expected asymptotic
behavior.
The best result is obtained for ${\rm B}=0.39$
and $\rho_{\rm c} = 0.1571\pm 0.0002$ and the corresponding
scaling plot is shown in Fig.\,\ref{fig:rho_a_4d_02}.
As one can see our data are consistent with the 
assumption that the asymptotic scaling behavior
of the four dimensional model obeys Eq.\,(\ref{eq:order_par_dc}).

\begin{figure}[t]
 \epsfxsize=7.1cm
 \epsffile{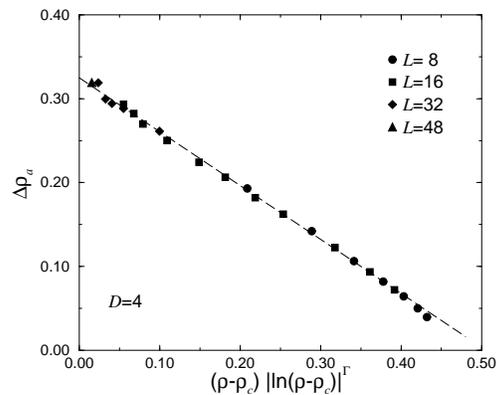}
  \caption{
     The fluctuations of the order parameter $\Delta \rho_{\rm a}$
     as a function of 
     $(\rho-\rho_{\rm c}) | \ln{(\rho-\rho_{\rm c})}|^\Gamma$
     in $D=4$.
     [see Eq.\,(\protect\ref{eq:fluc_01_dc})].
     Nearly straight lines are obtained for
     $\Gamma=1.66\pm0.1$.
     Thus the scaling behavior of the order parameter
     of the four dimensional model is governed by the
     mean-field exponent $\gamma=0$ modified by logarithmic
     corrections.
     The dashed line corresponds to a linear fit with
     the slope $-0.643$ and the intercept 
     $\Delta \rho_{\rm a}^{\scriptscriptstyle (0)}=0.325$.
   }
 \label{fig:rho_a_4d_04} 
\end{figure}

The mean-field behavior of the fluctuations is 
characterized by $\gamma=0$ which corresponds
to a jump~\cite{LUEB_20}.
Taking the logarithmic corrections into account
we assume that the asymptotic scaling behavior
of the fluctuations obeys the ansatz
\begin{equation}
\Delta \rho_{\rm a} \; = \; 
\Delta \rho_{\rm a}^{\scriptscriptstyle (0)}  \, - \,
{\rm const}\,
(\rho - \rho_{\rm c}) \,
| \ln{(\rho-\rho_{\rm c})}|^\Gamma .
\label{eq:fluc_01_dc}
\end{equation}
Plotting the fluctuations as a function of 
$(\rho - \rho_{\rm c}) \,| \ln{(\rho-\rho_{\rm c})}|^\Gamma$
one varies the correction exponent $\Gamma$ 
until one gets a straight line.
Convincing results are observed for $\Gamma=1.66\pm0.1$
and the corresponding plot is shown in 
Fig.\,\ref{fig:rho_a_4d_04}
(for $\Gamma>1.76$ and $\Gamma<1.56$ 
we observe significant curvatures).
The extrapolation to the vertical axis yield
the fluctuations at the transition point 
$\Delta \rho_{\rm a}^{\scriptscriptstyle (0)}=0.325\pm0.005$.

Thus we get that the scaling behavior of the 
four dimensional CLG model is characterized by the mean-field
exponents modified by logarithmic corrections
and we conclude therefore that the upper critical
dimension of the CLG model is $D_{\rm c}=4$.
This value agrees with the conjecture of a field
theory~\cite{PASTOR_2}.

\section{d=5}
\label{sec:d5}

\begin{figure}[t]
 \epsfxsize=7.1cm
 \epsffile{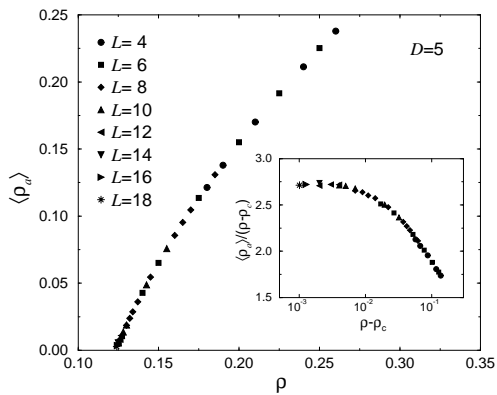}
  \caption{
    The average density of active sites $\langle \rho_{\rm a} \rangle$
    as a function of the density $\rho$ for various system sizes~$L$
    in $D=5$.
    The inset displays $\langle \rho_{\rm a} \rangle/(\rho-\rho_{\rm c})$
    as a function of $\rho-\rho_{\rm c}$.
    The saturation for $\rho-\rho_{\rm c}\to 0$ indicates that the
    asymptotic behavior of the order parameter agrees with the
    mean field behavior $\langle \rho_{\rm a} \rangle\sim\rho-\rho_{\rm c}$.
    The critical density is $\rho_{\rm c}=0.1230 \pm0.0004$.
   }
 \label{fig:rho_a_5d_01} 
\end{figure}

In the case of the five dimensional model
we considered system sizes from $L=4$ up to
$L=18$.
In the latter case $10^7$ update steps were used to 
reach the steady state close to the transition point.
The obtained values of the order parameter are 
plotted in Fig.~\ref{fig:rho_a_5d_01}.
The average density of active sites
seems to vanish linearly at the transition point.
This is supported by the logarithmic derivative 
[Eq.\,(\ref{eq:log_derivative})] which is plotted in
Fig.\,\ref{fig:log_diff}.
The effective exponent saturates in the vicinity of
the mean-field solution $\beta=1$.
One has to admit that the small available system sizes
in $d=5$ ($L\le 18$) hinder an analysis of the pure
critical scaling behavior.
But as we will see the data are sufficient to indicate
that the asymptotic scaling behavior is given
by $\beta=1$.
Therefore, we plot in the inset
of Fig.\,\ref{fig:rho_a_5d_01} 
$\langle \rho_{\rm a} \rangle/(\rho-\rho_{\rm c})$ as
a function of $\rho-\rho_{\rm c}$.
Approaching the transition point the rescaled
density of active sites indeed saturates.
Thus the five dimensional CLG model
is characterized by the mean-field scaling behavior 
\begin{equation}
\langle \rho_{\rm a} \rangle \; \sim \; 
\rho-\rho_{\rm c}.
\label{eq:rho_a_mean_field}
\end{equation} 

\begin{figure}[b]
 \epsfxsize=7.1cm
 \epsffile{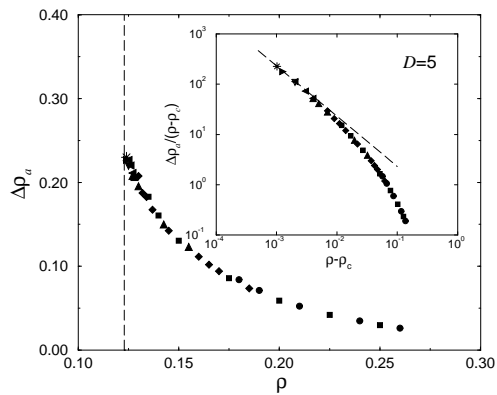}
  \caption{
    The fluctuations of the order parameter $\Delta \rho_{\rm a}$
    as a function of the density~$\rho$ in $D=5$.
    The symbols mark different system sizes~$L$
    (see Fig.\,\protect\ref{fig:rho_a_5d_01}).
    The dashed line marks the value of the critical density.
    The inset displays $\Delta \rho_{\rm a} /(\rho-\rho_{\rm c})$
    as a function of $\rho-\rho_{\rm c}$.
    The asymptotic behavior is characterized by a simple
    $(\rho-\rho_{\rm c})^{-1}$ dependence (dashed line) 
    which corresponds to the mean-field behavior (see text).
   }
 \label{fig:rho_a_5d_03} 
\end{figure}

The fluctuations of the order parameter $\Delta \rho_{\rm a}$
are plotted in Fig.\,\ref{fig:rho_a_5d_03}.
As one can see the fluctuations are characterized
by a jump at the transition point.
Following the mean-field behavior one expects
that the asymptotic behavior of the 
fluctuations is given by
\begin{equation}
\Delta \rho_{\rm a} \; = \; 
\Delta \rho_{\rm a}^{\scriptscriptstyle (0)} \, - \, 
{\rm const} \, (\rho - \rho_{\rm c}).
\label{eq:fluc_01_mf}
\end{equation}
In order to confirm this ansatz
we plot $\Delta \rho_{\rm a}/(\rho-\rho_{\rm c})$
as a function of $\rho-\rho_{\rm c}$.
If the above ansatz is valid the corresponding
curves have to display an asymptotic power-law behavior 
with the exponent $-1$ and the prefactor 
$\Delta \rho_{\rm a}^{\scriptscriptstyle (0)}$.
This behavior is indeed observed 
(see inset of Fig.\,\ref{fig:rho_a_5d_03})
and we get
$\Delta \rho_{\rm a}^{\scriptscriptstyle (0)}=0.231\pm0.006$.

Thus the asymptotic behavior of the
five dimensional CLG model is characterized by the
mean-field exponents.
This behavior strongly supports the above conclusion
that the upper critical dimension of the CLG model is
four.

\begin{table}[t]
\caption{The critical density $\rho_{\rm c}$ and 
the critical exponents $\beta$ and $\gamma$
of the CLG model for various dimensions~$D$.
The symbol $^{\ast}$ denotes logarithmic corrections
to the power-law behavior.
In the case of the three dimensional model the 
data of the fluctuation could be interpreted
as a small exponent or as a logarithmic growth
(see text).}
\label{table:exponents}
\begin{tabular}{llll}
$D$       &  $\rho_{\rm c}$	& $\beta$	& $\gamma$ \\  
\tableline \\
$2$     &  $0.34494\pm0.00003$	& $0.637\pm0.009$	& $0.384\pm0.023$   \\ 
$3$     &  $0.2179 \pm0.0001$	& $0.837\pm0.015$ & $0$ or $0.18\pm 0.06$  \\ 
$4$     &  $0.1571 \pm0.0002$	& $1^{\ast}$      & $0^{\ast}$ \\ 
$5$ 	&  $0.1230 \pm0.0004$	& $1$             & $0$ \\      
\end{tabular}
\end{table}

\section{Conclusions}
\label{sec:conc}

We analyse numerically the critical behavior of a
CLG model in various dimensions.
The values of the critical density 
and of the critical exponents are determined 
and the results are listed in Table~\ref{table:exponents}.
Our analysis suggests that four is 
the upper critical dimension of the
CLG model, i.e., the critical exponents 
depend on the value of the dimension 
for $D\le 4$.
Above this value we observe a mean-field
like scaling behavior.

\acknowledgments
I would like to thank A.~Hucht, H.\,K.~Janssen 
and A.~Vespignani for helpful discussions and 
useful comments on the manuscript.

\vspace{-0.3cm}

\end{document}